\documentclass[aps,prl,twocolumn,superscriptaddress,showpacs]{revtex4}

\usepackage{graphicx}
\usepackage{bm}
\usepackage{amsmath}
\usepackage{textcomp}
\usepackage{color}
\usepackage{ulem}

\begin{document}

\title{Towards Realizing a Quantum Memory for a Superconducting Qubit: 
Storage and Retrieval of quantum states}

\author{Shiro Saito}\email{saito.shiro@lab.ntt.co.jp}
\affiliation{NTT Basic Research Laboratories, NTT Corporation, 3-1, Morinosato-Wakamiya, Atsugi, Kanagawa 243-0198, Japan}

\author{Xiaobo Zhu}
\affiliation{NTT Basic Research Laboratories, NTT Corporation, 3-1, Morinosato-Wakamiya, Atsugi, Kanagawa 243-0198, Japan}

\author{Robert Ams\"{u}ss}
\affiliation{NTT Basic Research Laboratories, NTT Corporation, 3-1, Morinosato-Wakamiya, Atsugi, Kanagawa 243-0198, Japan}
\affiliation{Vienna Center for Quantum Science and Technology, Atominstitut, TU Wien, 1020 Vienna, Austria}

\author{Yuichiro Matsuzaki}
\affiliation{NTT Basic Research Laboratories, NTT Corporation, 3-1, Morinosato-Wakamiya, Atsugi, Kanagawa 243-0198, Japan}

\author{Kosuke Kakuyanagi}
\affiliation{NTT Basic Research Laboratories, NTT Corporation, 3-1, Morinosato-Wakamiya, Atsugi, Kanagawa 243-0198, Japan}

\author{Takaaki Shimo-Oka}
\affiliation{Graduate School of Engineering Science, Osaka University, 1-3 Machikane-yama, Toyonaka, Osaka 560-8531, Japan}

\author{Norikazu Mizuochi}
\affiliation{Graduate School of Engineering Science, Osaka University, 1-3 Machikane-yama, Toyonaka, Osaka 560-8531, Japan}

\author{Kae Nemoto}
\affiliation{National Institute of Informatics, 2-1-2 Hitotsubashi, Chiyoda-ku, Tokyo 101-8430, Japan}

\author{William J. Munro}
\affiliation{NTT Basic Research Laboratories, NTT Corporation, 3-1, Morinosato-Wakamiya, Atsugi, Kanagawa 243-0198, Japan}

\author{Kouichi Semba}
\affiliation{NTT Basic Research Laboratories, NTT Corporation, 3-1, Morinosato-Wakamiya, Atsugi, Kanagawa 243-0198, Japan}
\affiliation{National Institute of Informatics, 2-1-2 Hitotsubashi, Chiyoda-ku, Tokyo 101-8430, Japan}

\begin{abstract}

We have built a hybrid system composed of a superconducting flux qubit (the processor) and an ensemble of nitrogen-vacancy centers in diamond (the memory) that can be directly coupled to one another and demonstrated how information can be transferred from the flux qubit to the memory, stored and subsequently retrieved. We have established the coherence properties of the memory, and succeeded in creating an entangled state between the processor and memory, demonstrating how the entangled state's coherence is preserved. Our results are a significant step towards using an electron spin ensemble as a quantum memory for superconducting qubits.
\end{abstract}


\maketitle

Today's computers comprise a number of core components which all seemingly work together. As these devices are miniaturized, they are moving into the nanotechnology regime, where the quantum principles of superposition and entanglement can be used to process, store and transport information in radically new and powerful ways \cite{1,2}. Superconducting qubits have attracted a significant degree of interest in this context due to their ready fabrication, integratability, controllability and potential scalability \cite{4}. Quantum state manipulation and quantum algorithms have been successfully demonstrated on few qubit systems \cite{5}. However for larger scale realizations the currently reported coherence times of these macroscopic objects (superconducting qubits) \cite{14,15} has not yet reached those of microscopic systems (electron spins, nuclear spins $\dots$) \cite{6,7,8}. Hybridization between superconducting circuits and electron spin ensembles \cite{17,17-2,18,19,19-2,20,20-2,21,22,23} has been studied extensively as an alternative route to overcome this coherence limitation. Hybrid systems can potentially take the best elements of each individual system \cite{3}; the easy manipulation and processing of information in the superconducting circuits and the long storage times available in spin ensembles. This may allow the combined systems to have long enough coherence times for future large-scale quantum computation. The superconducting circuits can be used to process quantum information while the spin ensemble is used to store it. 

It is well known that the coupling strength of an individual electron spin to superconducting circuits is usually too small for the coherent exchange of quantum information. However, the coupling strength of an ensemble of $N$ spins can be enhanced by a factor of $\sqrt{N}$ \cite{24}. In the case of a superconducting resonator, $10^{12}$ electron spins are needed to achieve strong coupling \cite{19,20}. Kubo et al. \cite{21} have also succeeded in transferring a superposition state prepared in a transmon type qubit to a spin ensemble by way of the mm-size resonator.
On the contrary, a superconducting flux qubit can couple directly and more strongly to individual NV$^{-}$ centers and so only $10^{7}$ spins are needed to achieve strong coupling \cite{22,23}. Furthermore the sample occupies only a surface area of $\sim 50~\mu m^{2}$. This is highly attractive as many memories could be placed onto a single quantum based chip.

\begin{figure*}[htb]
\includegraphics[width=0.8\linewidth]{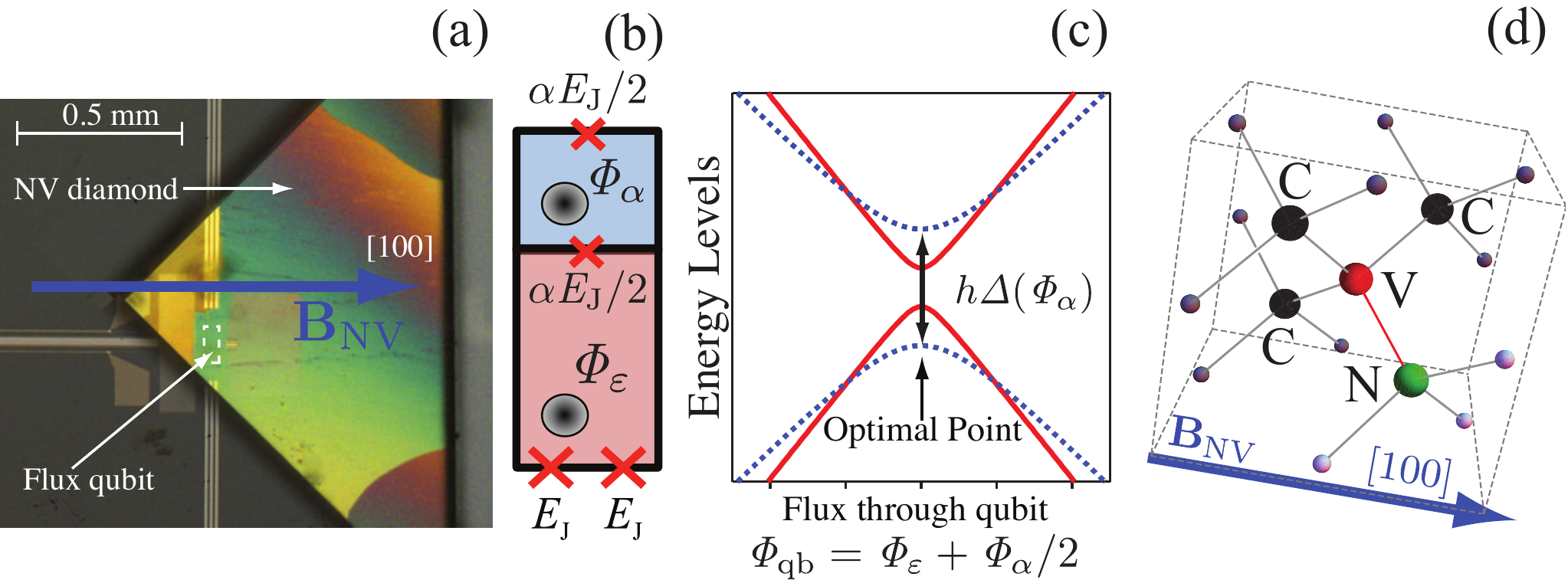}
\caption{
The flux qubit - NV$^{-}$ spin ensemble hybrid system. (a) Photograph of the hybrid system from the top. A diamond crystal is bonded on top of the flux qubit chip with its (001) surface facing the chip. An in-plane magnetic field $B_{\rm{NV}}$ is applied in the [100] crystallographic axis. (b) Schematic of the flux qubit. The qubit consists of the main loop (red, lower) and the $\alpha$-loop (blue, upper) and four Josephson junctions (red crosses) characterized by Josephson energies $E_{\rm{J}}$, $\alpha E_{\rm{J}}/2$. A typical $\alpha$ is about 0.9. $\varPhi_{\varepsilon}$ and $\varPhi_{\alpha}$ are the magnetic flux through the main loop and the $\alpha$-loop, respectively. Two broadband lines on the qubit chip control them. A dc-SQUID is attached to the qubit for readout of the qubit state (not shown) \cite{23}. (c) Energy levels of the qubit as a function of a qubit flux $\varPhi_{\rm{qb}}=\varPhi_{\varepsilon}+\varPhi_{\alpha}/2$. 
The energy imbalance $h \varepsilon = 2 I_{\rm{p}} \varPhi_{0} (\varPhi_{\rm{qb}}/\varPhi_{0} -1.5)$ between the two $I_{\rm{p}}$ states can be controlled by $\varPhi_{\rm{qb}} = \varPhi_{\varepsilon} + \varPhi_{\alpha}/2$. 
The energy splitting at the optimal point $h\varDelta$ can be controlled solely by $\varPhi_{\alpha}$ as shown by red curves and blue dotted curves. This enables us to couple the qubit to the NV$^{-}$ center at the optimal point where the qubit has its longest coherence time. (d) Structure of an NV$^{-}$ center in the diamond lattice. The red sphere is the vacancy while the green sphere is the nitrogen atom. Black spheres are the three nearest neighbor carbon atoms. The NV axis can be along four equivalent $\langle111\rangle$ directions, making the same angle of $55^\circ$ with $B_{\rm{NV}}$ parallel to [100]. 
\label{Fig1}}
\end{figure*}

Our hybrid system (depicted in Fig.~1(a)) consists of a diamond crystal incorporating dense NV$^{-}$ centers bonded on top of a gap tunable superconducting flux qubit \cite{25,26}. The ensemble was created by ion implantation of $^{12}\rm{C}^{2+}$ at 700 keV into type Ib (001) diamond with a nominal nitrogen concentration of 100 ppm followed by annealing in vacuum. This gave an NV$^{-}$ center concentration of $4.7 \times 10^{17}$ cm$^{-3}$ over approximately a 1 $\mu$m depth. The centers have an electron spin (S=1), with a zero field splitting $D$ = 2.88 GHz between the $|m_{\rm{s}} = 0 \rangle$ and $|m_{\rm{s}} = \pm1 \rangle$ levels at zero magnetic field. The spin Hamiltonian for a single NV$^{-}$ center is given by $H_{\rm{NV}}=h D S_{z}^{2} + h E (S_{x}^{2}-S_{y}^{2})+h g_{e} \mu_{\rm{B}} \bf{B}_{\rm{NV}}$ where $S_{x}$, $S_{y}$ and $S_{z}$ are the spin-one Pauli operators of $\bf{S}$; $E$ is the strain induced splitting; $h$ is Planck's constant; $g_{\rm{e}}$ = 2 is the NV$^{-}$ Land\'e factor; and $\mu_{\rm{B}}$ = 14 MHz/mT is the Bohr's magneton. In our experiment, we apply a static in-plane magnetic field $B_{\rm{NV}}$ to the diamond along the [100] crystalline axis to lift the degeneracy between $| \pm1 \rangle$ levels by approximately 80 MHz (Fig.~1(a),(d)). This reduces the strain $E$ effect on the decoherence. With the external magnetic field we can now store quantum states in a spin ensemble and evaluate its coherence which we could not achieve previously \cite{23}. The gap tunable flux qubit has a high nonlinearity and works as an effective two level system \cite{27} described by $H_{\rm{qb}}=\frac{h}{2} \left[\varepsilon (\varPhi_{\rm{qb}}) \sigma_{z} + \varDelta (\varPhi_{\alpha}) \sigma_{x } \right]$ where $\sigma_{z,x}$ are the usual Pauli spin operators (Fig.~1(c)). The eigenstates of $\sigma_{z}$ describe clockwise and counter-clockwise persistent current $I_{\rm{p}}$ states in the qubit and $h\varDelta$ represents the tunneling between them. 
The direct magnetic coupling between the qubit and the NV$^{-}$ spin ensemble is described by the interaction Hamiltonian \cite{22,23}
\begin{equation}
H_{\rm{int}}=h \sum_{k} \frac{g_{k}}{\sqrt{2}} \sigma_{+} \left( |0\rangle_{k} \langle +1|_{k} + |0\rangle_{k} \langle -1|_{k} \right) + H. c.
\end{equation}
where the sum runs over all the NV$^{-}$ centers coupled to the flux qubit; $g_k$ is the coupling strength of the qubit to the $k$-th NV$^{-}$ center; $|m \rangle_{k}$ is the $|m_{\rm{s}}=m\rangle$ state of the $k$-th center; $\sigma_{+} = |1 \rangle_{\rm{qb}} \langle 0|_{\rm{qb}}$ with $|0 \rangle_{\rm{qb}}$ and $|1 \rangle_{\rm{qb}}$ being the ground and excited states of the qubit, respectively. The collective coupling strength of the qubit to the spin ensemble of the $|m_{\rm{s}} = \pm 1 \rangle$ state can be represented by $g_{\rm{ens}}^{\pm} = \sqrt{\Sigma_{k} |g_{k}|^2/2} \sim\sqrt{N/2} \Bar{g}$, respectively. Here $\Bar{g} \sim 4.4$~kHz is the nominal coupling strength between the qubit and an NV$^{-}$ center, which was estimated from geometrical considerations \cite{23}. 
\begin{figure*}[htb]
\includegraphics[width=1.0\linewidth]{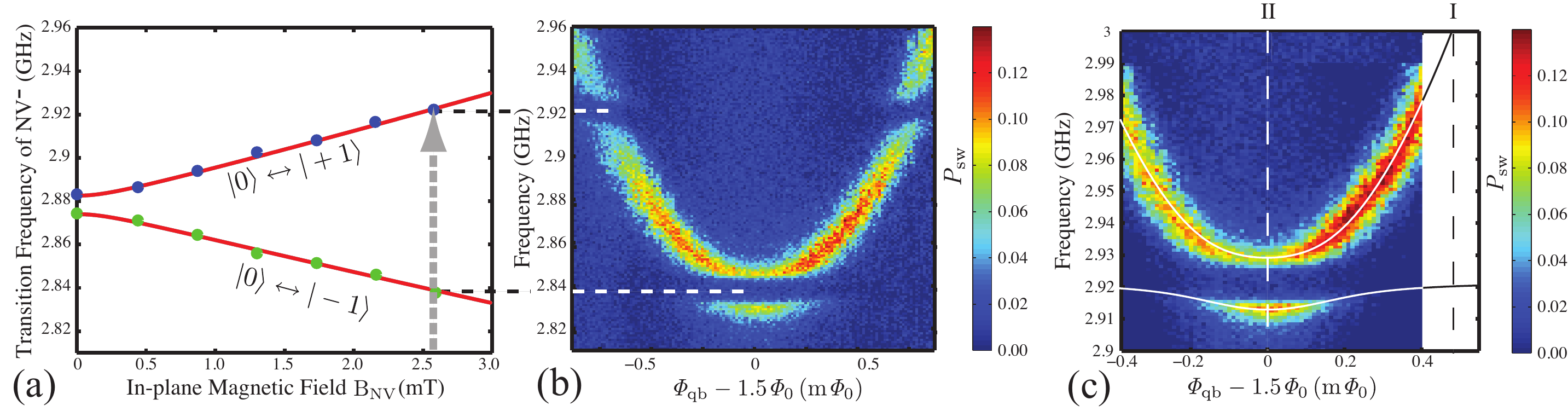}
\caption{
Energy spectrum of the flux qubit - NV$^{-}$ spin ensemble hybrid system. (a) Transition frequency of an NV$^{-}$ spin as a function of an in-plane magnetic field $B_{\rm{NV}}$. Dots represent the transition frequencies derived from the energy spectra of the flux qubit. 
Two curves are the results of the fitting to the Hamiltonian $H_{\rm{NV}}$ (see the main text). The parameters derived from the fitting are $D = 2.878$~GHz and $E = 4$~MHz. 
(b) Energy spectrum of the flux qubit coupled to the NV$^{-}$ spin ensemble under $B_{\rm{NV}}$ of 2.6~mT when the qubit splitting is tuned to $\varDelta = 2.84$~GHz. Here $P_{\rm{sw}}$ is proportional to the population of the qubit excited state \cite{note1}. In the series of experiments, the qubit ground (excite) state leads $P_{\rm{sw}}$ = 0 (0.45). 
(c) Energy spectrum of the flux qubit coupled to the NV$^{-}$ spin ensemble when $B_{\rm{NV}} = 2.6$~mT and $\varDelta = 2.92$~GHz. White curves represent transition frequencies of the hybrid system. The qubit is on resonance with the NV$^{-}$ spin ensemble at point II where we can perform an $i$SWAP operation based on a vacuum Rabi oscillation \cite{23}. 
Changing $\varPhi_{\rm{qb}}$ away from 1.5 $\varPhi_0$ increases the detuning between them resulting in the qubit being well decoupled from the spin ensemble at point I, where we perform qubit preparation, qubit detuning to store information in the memory and qubit readout. 
\label{Fig2}}
\end{figure*}

We started our experiments with the characterization of our NV$^{-}$ electron-spin ensemble under an in-plane magnetic field $B_{\rm{NV}}$. The measurements were performed in a dilution refrigerator at a mixing chamber temperature of 20 mK. The spectrum of the flux qubit under $B_{\rm{NV}}$ shows two energy anti-crossings when the qubit frequency is resonant with the NV$^{-}$ transitions $|0 \rangle_{k} \leftrightarrow |+1 \rangle_{k}$ and $|0 \rangle_{k} \leftrightarrow |-1 \rangle_{k}$ (Fig.~2(b)). We measured a number of spectra while changing $B_{\rm{NV}}$ and the NV$^{-}$ transition frequencies as a function of $B_{\rm{NV}}$ are well reproduced by the spin Hamiltonian $H_{\rm{NV}}$ (Fig.~2(a)). Then we tuned the qubit gap frequency $\varDelta$ to $f_{+} = 2.92$~GHz to utilize the spin ensemble of $m_{\rm{s}} = +1$ as a memory while applying a $B_{\rm{NV}} = 2.6$~mT field (Fig.~2(c)). The coupling strength of this transition $g_{\rm{ens}}^{+} = 8.9$ MHz derived from the spectrum indicates that approximately 10$^7$ NV$^{-}$ centers are collectively coupled to the flux qubit. 

The first step to demonstrate the memory capabilities of NV$^{-}$ centers is the storage of a single quantum excitation in the spin ensemble. The pulse sequence, depicted in the inset of Fig.~3(a), shows how to transfer the qubits excited state $|1 \rangle_{\rm{qb}} |0 \rangle_{\rm{ens}}$ to the memory $|0 \rangle_{\rm{qb}} |1 \rangle_{\rm{ens}}$, store it there for a time $\tau$ and then retrieve it. Here $|0 \rangle_{\rm{ens}} = |00 \dots 0 \rangle$ denotes the ground state of the spin ensemble with $|1 \rangle_{\rm{ens}} = (1/g_{\rm{ens}}^{+}) \Sigma_{k} \frac{g_k}{\sqrt{2}} S_{+,k} |0 \rangle_{\rm{ens}}$ being the first excited Dicke state of the spin ensemble with a single excitation in it where $S_{+,k} = |+1 \rangle_{k} \langle 0|_{k}$. Figure~3(a) shows the monotonic decay curve as the qubit is far detuned from the spin ensemble \cite{note2}. 
This means the information is successfully stored in the memory independently of the qubit. The memory time of the excited state is estimated to be 20.8$\pm$0.7~ns by fitting a simple exponential to the data. We should note that this procedure is usually used to measure an energy relaxation time $T_{\rm{1,ens}}$ of the collective mode of the spin ensemble. However the measured decay time is too short compared to that previously reported \cite{20}. This is because decoherence of the spin ensemble during its storage time is dominated by the inhomogeneous broadening probably due to a large electron spin-half bath derived from nitrogen impurities (P1 centers) \cite{21}, which changes the quantum state $|1 \rangle_{\rm{ens}}$ to $|1_{\theta} \rangle_{\rm{ens}} = (1/g_{\rm{ens}}^{+} ) \Sigma_{k}\frac{g_k}{\sqrt{2}} S_{+,k} e^{i \theta_k} |00 \dots 0 \rangle$ where $\theta_k$ is a phase shift associated with this broadening. The flux qubit can collectively couple to $|1 \rangle_{\rm{ens}}$ but not to $|1_{\theta} \rangle_{\rm{ens}}$. Hence the excitation cannot be transferred from the spin ensemble back to the qubit after the second $i$SWAP operation and so from the viewpoint of the qubit, this looks like relaxation from the memories excited state to another subspace. Hence we denote this decay time as an effective relaxation time $T_{\rm{1,ens}}^{*}$ of the spin ensemble. We should note that a numerical simulation can reproduce the data in Fig.~3(a) very well. From the simulation, we obtained an inhomogeneous broadening of 4.4~MHz with the Lorentz distribution and a hyperfine coupling to a nitrogen nuclear spin of 2.3~MHz, which reproduce our optically detected magnetic resonance (ODMR) spectrum of NV$^{-}$ centers at room temperature.

\begin{figure*}[htb]
\includegraphics[width=1.0\linewidth]{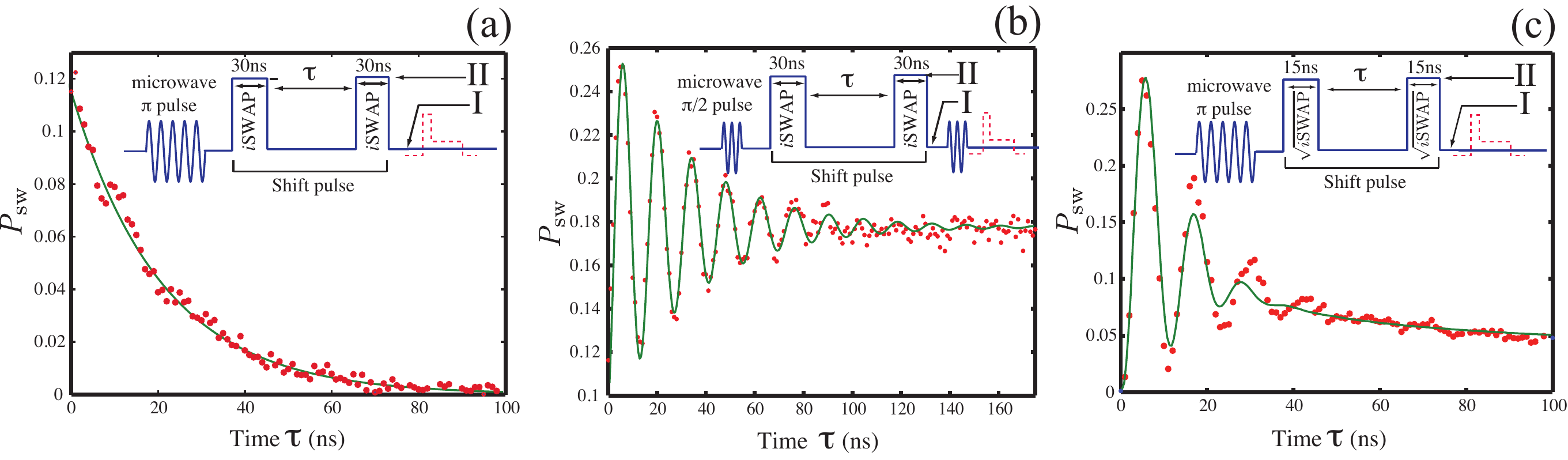}
\caption{
Quantum memory operations. The inset in each figure shows the pulse sequence used in the operations, which is applied to the qubit through the on-chip control line for $\varPhi_{\varepsilon}$. I and II represent the operating points indicated in Fig. 2(b). (a) Measurement results for the memory operation using a single quantum of energy. Red dots are the measured data with a green curve showing a fitted exponential decay with a decay time of $20.8\pm0.7$~ns (95\% confidence interval). The memory operation begins by first preparing the qubit in the excited state $|1\rangle_{\rm{qb}}$ using a microwave $\pi$ pulse (see the inset). An $i$SWAP gate transfers this excitation to the spin ensemble $|1\rangle_{\rm{ens}}$ and the qubit is then detuned. The $i$SWAP gate consists of a shift pulse tuning the qubit on resonance with the spin ensemble and its length is half of the period of a vacuum Rabi oscillation between them.
The detuning between the qubit and the spin ensemble during its storage time $\tau$ is approximately 80~MHz. Finally, the information (the excitation) is retrieved back from the spin ensemble to the qubit using a second $i$SWAP gate followed by its measurement using a readout SQUID (the red dashed pulse indicates a current pulse applied to the SQUID for the readout). (b) Measurement results for the memory operation using the superposition state $(|0\rangle_{\rm{ens}}+i |1\rangle_{\rm{ens}} )/\sqrt{2}$ with a detuning of approximately 70~MHz. Red dots are the measured data with the fitted green curve showing a decayed sinusoidal with a decay time of $33.6\pm2.3$~ns and a frequency of $70.4\pm0.3$~MHz (95\% confidence interval). (c) Storing one qubit of an entangled state in the spin ensemble quantum memory. Red dots are the measured data with the green curve showing a fitting to Eq. (2). Fitting parameters are $\zeta$ = 0.34, $\delta \omega$ = 87 MHz, $\phi$ = 0.85 radians.
\label{Fig3}}
\end{figure*}

The next step is to store an arbitrary superposition state which we choose for convenience as $(|0\rangle_{\rm{ens}}+i |1\rangle_{\rm{ens}})/\sqrt{2}$. The inset of Fig.~3(b) depicts the pulse sequence used for this experiment. The sequence corresponds to a Ramsey fringe experiment for the spin ensemble and enables us to estimate the spin ensemble's dephasing time $T_{\rm{2,ens}}^{*}$. A fitting of an exponential decaying sinusoidal to the data in Fig.~3(b) yields $T_{\rm{2,ens}}^{*}=33.6\pm2.3$~ns (similar to \cite{28}). We can thus estimate a pure dephasing time $T_{\rm{2,ens}}$ to be 175~ns from the relation $(T_{\rm{2,ens}}^{*} )^{-1}=(2T_{\rm{1,ens}}^{*} )^{-1}+(T_{\rm{2,ens}} )^{-1}$. The physical origin of the pure dephasing is unclear but P1 centers far from our NV$^{-}$ centers layer and charge effects on the diamond surface caused by the ion implantation and subsequent treatment can be possibilities \cite{note3}. 

The storage of single qubits states is an excellent start but we also need to be able to work with entangled states. Hence we create a two qubit Bell state and show our memory can maintain coherence within this two-qubit state. For this experiment, we adopted a pulse sequence shown in the inset of Fig.~3(c). The first $\sqrt{i\rm{SWAP}}$ pulse creates an entangled state of the form $(i|0\rangle_{\rm{qb}} |1\rangle_{\rm{ens}} + |1\rangle_{\rm{qb}} |0\rangle_{\rm{ens}})/\sqrt{2}$ \cite{note4}. The flux qubit is moved off-resonance for a time $\tau$ allowing one qubit of the entangled state to be stored. A second $\sqrt{i\rm{SWAP}}$ (on resonance) then transfers the phase information of the entangled state to the qubits population, which is measured. 
Figure~3(c) shows the measurement outcome with a large detuning $\delta \omega$ during the storage so that the qubit and the spin ensemble are well decoupled. 
The oscillation reflects a phase evolution of the entangled state during the storage and it can be qualitatively reproduced by 
\begin{eqnarray}
P_{\rm sw} &=& \zeta \bigr[ e^{- {\it \tau}/T_{1,{\rm qb}}} \cos^4 \phi +  e^{- {\it \tau}/T^*_{1,{\rm ens}}} \sin^4 \phi  \nonumber \\ 
&-& 2 e^{-\Lambda}  \sin^2 \phi \cos^2 \phi \cos \left(\delta \omega t \right) \bigl]
\end{eqnarray}
where $\Lambda=\frac{\it \tau}{2T_{\rm 1,qb}} + \frac{\it \tau}{2T^*_{\rm 1,ens}} + (\frac{\it \tau}{T^{*}_{\rm 2,qb}})^2 +\frac{\it \tau}{T_{\rm 2,ens}}$, $\zeta$ is a scaling parameter and $\phi$ represents imperfection of $\sqrt{i\textrm{SWAP}}$ pulses \cite{note8}. From the previous memory experiments we have established $T^*_{1,{\rm ens}}\sim 20.8$ ns, $T_{2,{\rm ens}} \sim 175$ ns. Independent measurements on the flux qubit (under the same conditions) have also established $T_{1,{\rm qb}} \sim 395$ ns and $T^{*}_{2,{\rm qb}}=19.7$ ns \cite{note5}. This only leave $\zeta, \delta \omega, \phi$ as a free parameter which we fit in Fig.~3(c). Interestingly our experimental results show longer oscillation than that derived from eq.~(2) with previously established coherence times, which could be explained by collective noise \cite{note7} and a decoherence free subspace \cite{29}.

The above experiments suggest that the memory times of the NV$^{-}$ spin ensemble are limited by inhomogeneous broadening probably due the P1 centers and the hyperfine coupling to the nitrogen nuclear spin. 
To further enhance the storage time, we need to reduce the density of P1 centers \cite{30}, polarize the nuclear spins and at the same time to increase the coupling between the flux qubit and a single NV$^{-}$ spin to maintain the same collective coupling strength. For instance increasing the persistent current $I_{\rm{p}}$ by a factor of ten and decreasing the distance between the qubit and the NV$^{-}$ spin to one-tenth (by fabricating the qubit directly on the diamond) means several hundred NV$^{-}$ spins should be enough to maintain the strong coupling. Also adopting a (110) diamond crystal and applying an in-plane magnetic field in the [111] direction (as shown by \cite{21}) will further enhance the coherence of the spin ensemble. Finally, by using a spin echo technique we could cancel out the affect of the inhomogeneous broadening and also decouple the nitrogen atoms nuclear spin from our electron spin further enhancing the ensembles coherence time and hence storage time \cite{31,32}.

In summary, we have demonstrated quantum memory operations in a superconductor-diamond (flux qubit - NV$^{-}$ spin ensemble) hybrid system. The quantum state prepared in a superconducting flux qubit, has been directly transferred to, stored and retrieved from an NV$^{-}$ spin ensemble in diamond. Furthermore we have stored one qubit of an entangled state in the memory, retrieving it later and shown the coherence is preserved. The direct and strong coupling between the flux qubit and the memory enables us to utilize a much smaller size ensemble than in similar resonator based experiment, which should in the future allow us to have inherently longer coherence times. This is a significant step towards the realization of a long-lived quantum memory for superconducting flux qubits. 

\begin{acknowledgments}
We thank S. Karimoto, H. Nakano, M. Kasu, H. Tanji, B. Pingault, J. Schmiedmayer, J. Mayers, M. S. Everitt, B. Scharfenberger and A. Kemp for valuable discussions during this project. This work was supported in part by KAKENHI (A) grant no. 22241025 and the FIRST program and NICT. R A. was supported by the Doctoral Programme CoQuS (W1210). 
\end{acknowledgments}


\begin{thebibliography}{}
\bibitem{note1} To obtain the spectrum, a 1~$\mu$s long microwave pulse is applied to the flux qubit followed immediately by a current pulse to the dc-SQUID for readout. If the qubit is detected as being in $|0\rangle_{\rm{qb}}$, the SQUID switches to a voltage state, otherwise it stays in a superconducting state. By repeating the measurement 4,000 times, we obtain the SQUID switching probability. We define $P_{\rm{sw}}$ as the subtraction of the switching probability from that without applying the microwave pulse to eliminate any offset switching probability. For the time domain measurements in Fig.~3, we repeat the measurement 20,000 times.
\bibitem{note2} We can perform the memory operations with various detuning during the storage, where the detuning is represented by the qubit flux $\varPhi_{\rm{qb}}$. In the range -0.3~m$\varPhi_{0} \le \varPhi_{\rm{qb}}-1.5$~$\varPhi_{0} \le 0.4$~m$\varPhi_{0}$, the detuning is not enough to decouple the qubit from the spin ensemble so that vacuum Rabi oscillations are observed. On the other hand, when the qubit is far detuned (more than 50~MHz) from the spin ensemble, namely 0.4~m$\varPhi_{0} < \varPhi_{\rm{qb}}-1.5$~$\varPhi_{0}$, monotonic decay curves are observed.
\bibitem{note3} We do not believe this dephasing is caused by fluctuations in the in-plane magnetic field $B_{\textrm{NV}}$. Our external coils used to generate $B_{\textrm{NV}}$ affects both the flux qubit and diamond sample. We would
 have observed very bad coherence properties for the flux qubit if this
 was the dephasing source. However the coherence properties of our flux qubit
 did not change whether $B_{\textrm{NV}}$ was  applied or not.
 Nor do we believe this pure dephasing is arising from the nitrogen
 nuclear spin present in every NV$^-$ center. The effect of an nuclear
 spin is quite localized to it neighbouring area and will not have an
 effect across the entire ensemble.
\bibitem{note4} This is a maximally entangled state where a single excitation is coherently shared between the qubits.
\bibitem{note8} This formula is derived in the following way. We begin with the flux qubit prepared in its excited state $|1\rangle_{\rm qb}$. Then an imperfect $\sqrt{i\rm{SWAP}}$ gate is performed to create the state $\cos \phi |1\rangle_{\rm qb} |0\rangle_{\rm ens} + i \sin \phi |0\rangle_{\rm qb} |1\rangle_{\rm ens}$. This means the perfect $\sqrt{i\rm{SWAP}}$ gate would give $\phi = \pi/4$. The qubit and ensembles are then assumed to be far off resonance with each other so that independent relaxation and dephasing processes can takes place. This allows us to include $T_{1,{\rm qb}}$, $T^{*}_{2,{\rm qb}}$,  $T^*_{1,{\rm ens}}$ and $T_{2,{\rm ens}}$ effects. A second imperfect $\sqrt{i\rm{SWAP}}$ gate can then be performed and the qubit state determined. The probability for the qubit being in its excited state is then given by eq.~(2).
\bibitem{note5} The qubit relaxation $T_{1,{\rm qb}}$ and coherence $T^{*}_{2,{\rm qb}}$ was measured at point I in Fig. 2(c) which is far from the optimal point. Typical $T^{*}_{2,{\rm qb}}$ at the optimal point is about 100 ns.
\bibitem{note7} We can not tell definitely where the collective noise is coming from. However charge effects on the diamond surface as well as P1 centers in the bulk (non NV${^-}$ region) should affect both the flux qubit and the spin ensemble.
\end{thebibliography}
\end{document}